\documentclass{amsart}
\usepackage{graphicx}
\usepackage{epic,eepic,eepicemu}
\usepackage{subfigure}
\usepackage[normalem]{ulem}
\usepackage{epsfig}

\usepackage[usenames,dvipsnames]{pstricks}
\usepackage{pst-grad} 
\usepackage{pst-plot} 

\newtheorem{thm}{Theorem}[section]

\newtheorem{lem}[thm]{Lemma}
\newtheorem{prop}[thm]{Proposition}
\newtheorem{prob}[thm]{Problem}
\newtheorem{conj}[thm]{Conjecture}
\newtheorem{Clm}{Claim}[thm]
\newtheorem{rem}[thm]{Remark}
\theoremstyle{definition}

\theoremstyle{remark}

\newenvironment{prf}{{\bf \noindent Proof } }{\hfill$\square$\\}
\newenvironment{PrfClaim}{{\bf Proof }}{{\hfill\tiny{$\square$\\}}}

\title[]{On Fulkerson conjecture}
\author{J.L. Fouquet and J.M. Vanherpe}
\address{L.I.F.O., Facult\'e des Sciences, B.P. 6759 \\
Universit\'e d'Orl\'eans, 45067 Orl\'eans Cedex 2, FR}

\subjclass{035 C} 

\begin{document}
\input{epsf.sty}
\begin{abstract}
If $G$ is a bridgeless cubic graph, Fulkerson conjectured that we
can find  $6$ perfect matchings (a {\em Fulkerson covering})  with
the property that every edge of $G$ is contained in exactly two of
them. A consequence of the Fulkerson conjecture would be that every
bridgeless cubic graph has $3$ perfect matchings with empty
intersection (this problem is known as the Fan Raspaud Conjecture).
A {\em FR-triple} is a set of $3$ such perfect matchings. We show
here how to derive a Fulkerson covering from two FR-triples.

Moreover, we give a simple proof that the Fulkerson conjecture holds
true for some classes of well known snarks.
\end{abstract}

\maketitle

\section{Introduction}
The following conjecture is due to Fulkerson, and appears first in
\cite{Ful71}.
\begin{conj}\label{Conjecture:Fulkerson} If $G$ is a bridgeless
cubic graph, then there exist $6$ perfect matchings $M_1,\ldots,M_6$
of $G$ with the property that every edge of $G$ is contained in
exactly two of $M_1,\ldots,M_6$.
\end{conj}

We shall say that  $\mathcal F=\{M_{1} \ldots M_{6}\}$, in the above
conjecture, is a {\em Fulkerson covering}. A consequence of the
Fulkerson conjecture would be that every bridgeless cubic graph has
$3$ perfect matchings with empty intersection (take any $3$ of the
$6$ perfect matchings given by the conjecture). The following
weakening of this (also suggested by Berge) is still open.

\begin{conj}\label{Conjecture:Berge2}There exists a fixed integer $ k $
such that every bridgeless cubic graph has a list of $ k $ perfect
matchings with empty intersection.
\end{conj}

For $k=3$ this conjecture is known as the Fan Raspaud Conjecture.

\begin{conj}\cite{FanRas} \label{Conjecture:FanRaspaud} Every
bridgeless cubic graph contains perfect matching $M_1$, $M_2$, $M_3$
such that
$$M_1 \cap M_2 \cap M_3 = \emptyset$$
\end{conj}

Let $G$ be a cubic graph with $3$ perfect matchings $M_{1}, M_{2}$
and $M_{3}$ having an empty intersection. Since $G$ satisfies the
Fan Raspaud conjecture, when considering these perfect matchings,
we shall say that $\mathcal T=(M_{1}, M_{2},M_{3})$ is a {\em
FR-triple}. We define $T_{i} \subset E(G)$ ($i=0..2$) as the set of
edges of $G$ which are covered $i$ times by $\mathcal T$. It will be
convenient to use $T'_{i}$($i=0..2$) for the FR-triple $\mathcal
T'$.

\section{FR-triples and Fulkerson covering}

 In
this section, we are concerned with the relationship between
FR-triples and Fulkerson coverings.

\subsection{On FR-triples}

\begin{prop} \label{Proposition:StructureFR}
Let $G$ be a bridgeless cubic graph with $\mathcal T$ a FR-triple.
Then $T_{0}$ and $T_{2}$ are disjoint matchings.
\end{prop}
\begin{prf}
Let $v$ be a vertex incident to an edge of $T_{0}$. Since $v$ must
be incident to each perfect matching of $\mathcal T$ and since the
three perfect matchings have an empty intersection, one of the
remaining edges incident to $v$ must be contained in $2$ perfect
matchings while the other is contained in exactly one perfect
matching. The result follows.
\end{prf}

\begin{figure}
\includegraphics{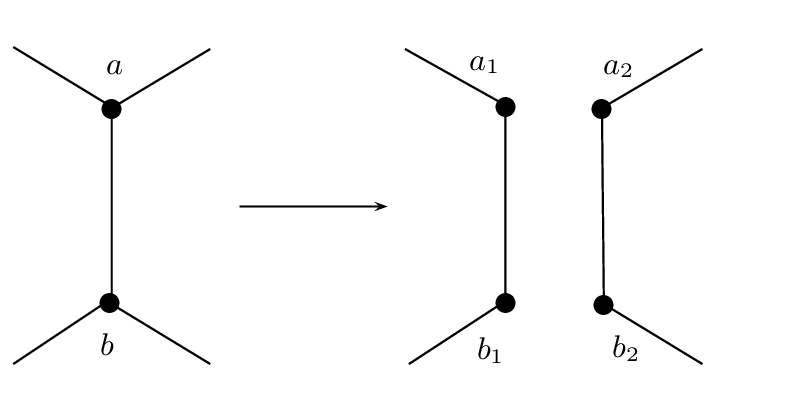}
\caption{ Splitting an edge}
\label{Figure:SplittingEdge}
\end{figure}

We introduce now  concepts and definitions coming from
\cite{HaoNiuWanZhaZha2009}. Let $ab$ be an edge of bridgeless cubic
graph $G$. We shall say that we have {\em splitted} the edge $ab$
when we have applied the operation depicted in Figure
\ref{Figure:SplittingEdge}. The resulting graph is no longer cubic
since we get $4$ vertices with degree $2$ instead of two vertices of
degree $3$. Let $A_{1}$ and $A_{2}$ be two disjoint matchings of $G$
(we insist to say that these matchings are not, necessarily, perfect
matchings).  For $i=1,2$, let $G_{A_{i}}$ be the graph obtained by
splitting the edges of $A_{i}$ and let $\overline{G_{A_{i}}}$ be the
graph homeomorphic to $G_{A_{i}}$ when the degree $2$ vertices are
deleted. The connected component of $\overline{G_{A_{i}}}$ are cubic
graphs and {\em vertexless loop graphs} (graph with one edge and no
vertex). We shall say that $\overline{G_{A_{i}}}$  is $3-$edge
colourable whenever the cubic components are $3-$edge colourable
(any colour can be given to the vertexless loops).

The following Lemma can be obtained from the work of Hao and
al. \cite{HaoNiuWanZhaZha2009} when considering FR-triples.

\begin{lem} \label{Lemma:FRGBarreT2}
Let $G$ be a bridgeless cubic graph and let $\mathcal T$ be a
FR-triple. Then $\overline{G_{T_{2}}}$ is $3-$edge colourable.
\end{lem}
\begin{prf}

Assume that $T=(M_{1}, M_{2},M_{3})$ is a FR-triple.  Let $ab$ be an
edge of $T_{2}$ then the two edges of $T_{1}$ incident with $ab$
must be in the same perfect matching of $\mathcal T$. Hence, these
two edges are identified in some sens.  If we colour the edges of
$T_{1}$ with $1$, $2$ or $3$ when they are in $M_{1}$, $M_{2}$ or
$M_{3}$ respectively, we get a natural $3-$edge colouring of
$\overline{G_{T_{2}}}$.
\end{prf}

\begin{lem} \label{Lemma:GBarreA1FR}
Let $G$ be a bridgeless cubic graph containing  two disjoint
matchings $A_{1}$ and $A_{2}$ such that $\overline{G_{A_{1}}}$ is
$3-$edge colourable and $A_{1} \cup A_{2}$ forms an union of disjoint cycles. Then $G$ has a FR-triple $\mathcal T$  where
$T_{2}=A_{1}$ and $T_{0}=A_{2}$.
\end{lem}
\begin{prf}
Obviously,$A_{1} \cup A_{2}$ forms an union of disjoint even cycles
in $G$. Let $C=a_{0}a_{1} \ldots a_{2p-1}$ be an even cycle of
$A_{1} \cup A_{2}$ and assume that $a_{i}a_{i+1} \in A_{1}$ when $i
\equiv 0 (2)$.

Let $M_{1}$, $M_{2}$ and  $M_{3}$ be the three matchings associated
to a $3$ edge-colouring of $\overline{G_{A_1}}$.  Thanks to the
construction of $\overline{G_{A_1}}$ for some $i\equiv 0[2]$, the
third edge incident to $a_i$, say $e$, and the third one incident to
$a_{i+1}$, say $e'$ lead to a unique  edge of $\overline{G_{A_1}}$.
Assume that this edge of $\overline{G_{A_1}}$ is in $M_{1}$, then
$M_{1}$ can be extended naturally to a matching of $G$ containing
$\{e,e'\}$. Moreover we add $a_ia_{i+1}$ to $M_{2}$ and $a_ia_{i+1}$
to $M_{3}$. When applying this process to all edges of $A_1$ on all
cycles of $A_1\cup A_2$ we extend the colours of
$\overline{G_{A_1}}$ into perfect matchings of $G$. Since every edge
of $G$ belongs to at most $2$ matchings in $\{M_{1}, M_{2}, M_{3}\}$
we have a FR-triple with $\mathcal T=\{M_{1}, M_{2}, M_{3}\}$. By
construction, we have $T_{2}=A_{1}$ and $T_{0}=A_{2}$, as claimed.

\end{prf}

\begin{prop} \label{Proposition:EquivalenceStructureFR}
Let $G$ be a bridgeless cubic graph then $G$ has a FR-triple if and
only if $G$ has  two disjoint matchings $A_{1}$ and $A_{2}$ such
that $A_{1} \cup A_{2}$ forms an union of disjoint cycles, moreover $\overline{G_{A_{1}}}$ or $\overline{G_{A_{2}}}$ is $3-$edge
colourable.
\end{prop}
\begin{prf}
Assume that $G$ has  two disjoint matchings $A_{1}$ and $A_{2}$ such
that, without loss of generality, $\overline{G_{A_{1}}}$ is $3-$edge
colourable. From Lemma \ref{Lemma:GBarreA1FR}, $G$ has a FR-triple
$\mathcal T$ where $T_{2}=A_{1}$ and $T_{0}=A_{2}$.

Conversely, assume that $\mathcal T$ is a FR-triple. From Lemma
\ref{Lemma:FRGBarreT2} $\overline{G_{T_{2}}}$ is $3$-edge
colourable.  Let $A_{1}=T_{0}$ and $A_{2}=T_{2}$. Then $A_{1}$ and $
A_{2}$ are two disjoint matchings and $\overline{G_{A_{2}}}$ is
$3-$edge colourable.
\end{prf}

\subsection{On compatible FR-triples}
As pointed out in the introduction, any three perfect matchings in a
Fulkerson covering lead to a FR-triple. Is it possible to get a
Fulkerson covering when we know one or more FR-triples?   In fact,
we can characterize a Fulkerson covering in terms of FR-triples in
the following way.

Let $G$ be a bridgeless cubic graph with $\mathcal T=(M_{1},
M_{2},M_{3})$ and $\mathcal T'=(M'_{1}, M'_{2},M'_{3})$ two
FR-triples. We shall say that $\mathcal T$ and $\mathcal T'$ are
{\em compatible} whenever $T_{0}=T'_{2}$ and $T_{2}=T'_{0}$ (and
hence  $T_{1}=T'_{1}$).

\begin{thm}\label{Theorem:CompatibleFRTriples}
Let $G$ be a bridgeless cubic graph then $G$ can be provided with a
Fulkerson covering if and only if $G$ has  two compatible
FR-triples.
\end{thm}

\begin{prf}
Let $\mathcal F=\{M_{1} \ldots M_{6}\}$ be a Fulkerson covering of
$G$ and let $\mathcal T=(M_{1}, M_{2},M_{3})$ and $\mathcal
T'=(M_{4}, M_{5},M_{6})$. $\mathcal T$ and $\mathcal T'$ are two
FR-triples and we claim that they are compatible. Since each edge of
$G$  is covered exactly twice by $\mathcal F$, $T_{1}$ the set of
edges covered only once by $\mathcal T$ must be covered also only
once by $\mathcal T'$, $T_{0}$ the set of edges not covered by
$\mathcal T$ must be covered exactly twice by $\mathcal T'$ and
$T_{2}$ the set of edges covered exactly twice by $\mathcal T$ is
not covered by $\mathcal T'$. Which means that $T_{1}=T'_{1}$,
$T_{0}=T'_{2}$ and $T_{2}=T'_{0}$, that is $\mathcal T$ and
$\mathcal T'$ are compatible.

Conversely, assume that $\mathcal T$ and $\mathcal T'$ are two
FR-triples compatible. Then it is an easy task to check that each
edge of $G$ is contained in exactly $2$ perfect matchings of the $6$
perfect matchings involved in $\mathcal T$ or $\mathcal T'$.
\end{prf}

\begin{prop}\label{Proposition:EquivalenceFRTripleFulkerson}
Let $G$ be a bridgeless cubic graph then $G$ has two compatible
FR-triples if and only if $G$ has  two disjoint matchings $A_{1}$
and $A_{2}$ such that $A_{1} \cup A_{2}$ forms an union of disjoint cycles and $\overline{G_{A_{1}}}$ and
$\overline{G_{A_{2}}}$ are $3-$edge colourable.
\end{prop}
\begin{prf}
Let $\mathcal T$ and $\mathcal T'$ be $2$ compatible FR-triples.
From Lemma \ref{Lemma:FRGBarreT2} we know that
$\overline{G_{T_{2}}}$ and $\overline{G_{T^{'}_{2}}}$ are $3-$edge
colourable. Since $T_{0}=T^{'}_{2}$ and $T^{'}_{0}=T_{2}$ by the
compatibility of $\mathcal T$ and $\mathcal T^{'}$, the result holds
when we set $A_{1}=T_{0}$ and $A_{2}=T_{2}$.

Conversely, assume that $G$ has  two disjoint matchings $A_{1}$ and
$A_{2}$ such that $\overline{G_{A_{1}}}$ and $\overline{G_{A_{2}}}$
are $3-$edge colourable. From Lemma \ref{Lemma:GBarreA1FR}, $G$ has
a FR-triple $\mathcal T$ where $T_{2}=A_{1}$ and $T_{0}=A_{2}$ as
well as a FR-triple $\mathcal T^{'}$ where $T^{'}_{2}=A_{2}$ and
$T^{'}_{0}=A_{1}$. These two FR-triples are obviously compatible.
\end{prf}

\begin{prop}\cite{HaoNiuWanZhaZha2009}\label{Proposition:HaoNiuWanZhanZhan}
Let $G$ be a bridgeless cubic graph then $G$ can be provided with a
Fulkerson covering if and only if $G$ has  two disjoint matchings
$A_{1}$ and $A_{2}$ such that $A_{1} \cup A_{2}$ forms an union of disjoint cycles and $\overline{G_{A_{1}}}$ and
$\overline{G_{A_{2}}}$ are $3-$edge colourable.
\end{prop}
\begin{prf}
Obvious in view of Theorem \ref{Theorem:CompatibleFRTriples} and
Proposition \ref{Proposition:EquivalenceFRTripleFulkerson}.
\end{prf}

\section{Fulkerson covering for some classical snarks}
A non $3-$edge colourable, bridgeless, cyclically $4-$edge-connected
cubic graph is called a {\em snark}. 

\begin{figure}
\centering \epsfsize=0.45 \hsize \noindent \epsfbox{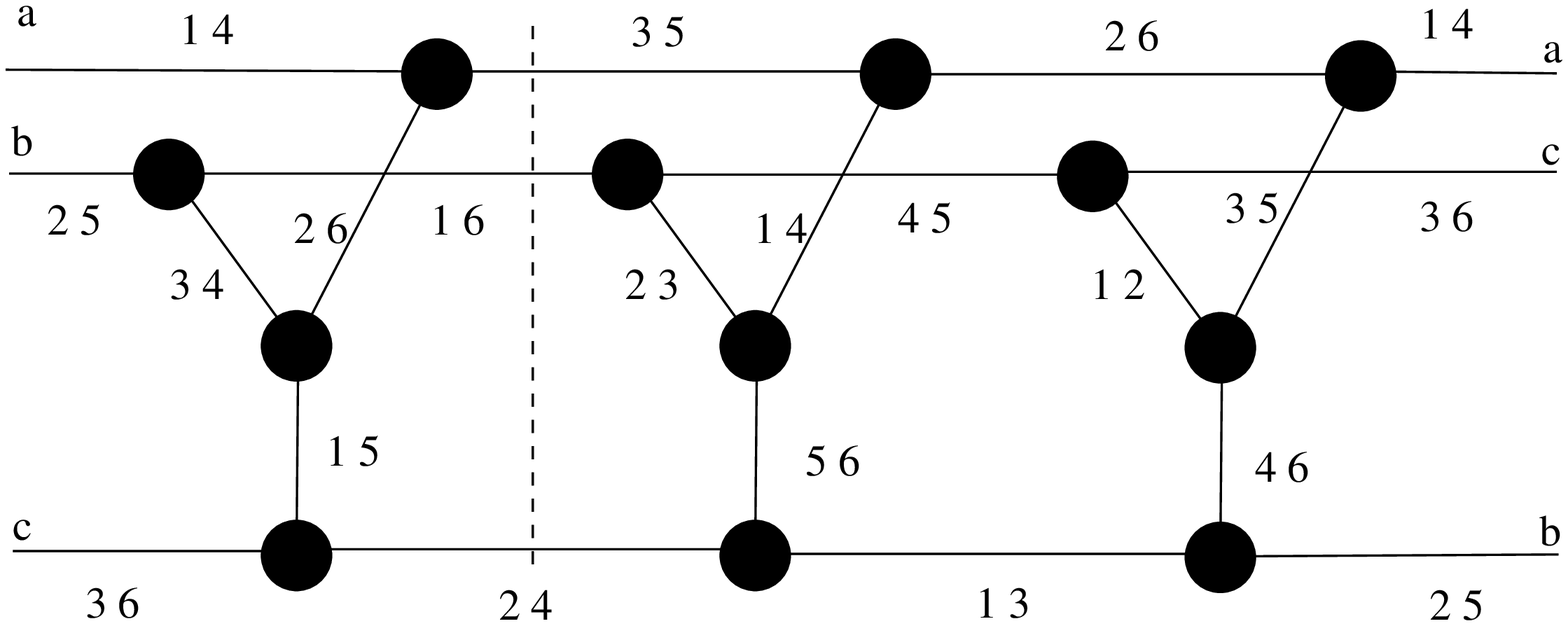}
\caption{$J_{3}$} \label{Figure:J3}
\end{figure}

For an odd $k\geq 3$, let $J_k$ be the cubic graph on $4k$ vertices $x_0,x_1,\ldots
x_{k-1}$, $y_0,y_1,\ldots y_{k-1}$, $z_0,z_1,\ldots z_{k-1}$,
$t_0,t_1,\ldots t_{k-1}$ such that $x_0x_1\ldots x_{k-1}$ is an
induced cycle of length $k$, $y_0y_1\ldots y_{k-1}$ $z_0z_1\ldots
z_{k-1}$ is an induced cycle of length $2k$ and for $i=0\ldots k-1$
the vertex $t_i$ is adjacent to $x_i$, $y_i$ and $z_i$. The set
$\{t_i,x_i,y_i,z_i\}$ induces the claw $C_i$. In Figure
\ref{Figure:J3} we have a representation of $J_{3}$, the half edges
(to the left and to the right in the figure) with same labels are
identified.
For $k\geq 5$ those graphs were introduced by Isaacs in \cite{Isa75} under the name of flower snarks in order to provide an infinite family of snarks.

Proposition \ref{Proposition:HaoNiuWanZhanZhan} is essentially used in
\cite{HaoNiuWanZhaZha2009} in order to show that the so called
flower snarks and Goldberg snarks can be provided with a Fulkerson
covering. We shall see, in this section, that this result can be
directly obtained.

\begin{thm}  \label{Theorem:FlowerSnark} For any odd $k\geq 3$, $J_{k}$ can be
provided with a Fulkerson covering.
\end{thm}
\begin{prf}
For $k=3$ the Fulkerson covering is given in Figure \ref{Figure:J3}.
We obtain a Fulkerson covering of $J_{k}$ by inserting a suitable
number of subgraphs isomorphic to the subgraph depicted in Figure
\ref{Figure:JMaillon} when we cut $J_{3}$ along the dashed line of
Figure \ref{Figure:J3}. The labels of the edges of the two sets of
three semi-edges (left and right) are identical which insures that
the process can be repeated as long as necessary. These labels lead
to the perfect matchings of the Fulkerson covering.

\begin{figure}
\includegraphics{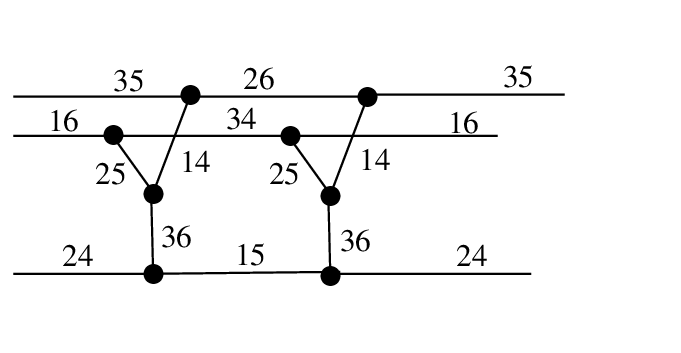}
\caption{A block for the flower snark} \label{Figure:JMaillon}
\end{figure}
\end{prf}

\begin{figure}
\includegraphics{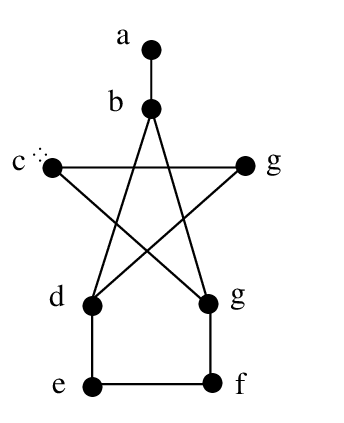}
\caption{$H$} \label{Figure:Hi}
\end{figure}

Let $H$ be the graph depicted in Figure \ref{Figure:Hi}

Let $G_{k}$ ($k$ odd) be a cubic graph obtained from $k$ copies of
$H$ ($H_{0} \ldots H_{k-1}$ where the name of vertices are indexed
by $i$) by adding edges  $a_{i}a_{i+1}$, $c_{i}c_{i+1}$,
$e_{i}e_{i+1}$, $f_{i}f_{i+1}$ and $h_{i}h_{i+1}$ (subscripts are
taken modulo $k$).

\begin{figure}
\centering 
\includegraphics[scale=0.8]{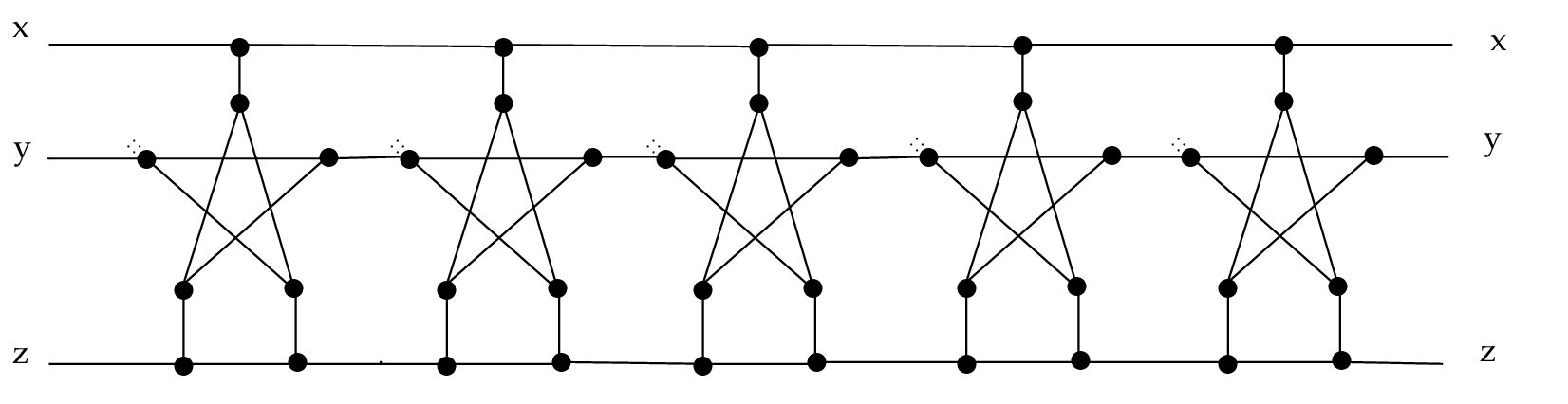}
\caption{Goldberg snark $G_5$} \label{Figure:Goldberg5}
\end{figure}

If $k = 5$, then $G_k$ is known as the Goldberg snark (see
\cite{Gol81}). Accordingly, we refer to all graphs $G_k$ as Goldberg
graphs. The graph $G_5$ is shown in Figure \ref{Figure:Goldberg5}.
The half edges (to the left and to the right in the figure) with
same labels are identified.

\begin{thm}  \label{Theorem:GoldbergSnark} For any odd $k\geq 5$, $G_{k}$ can be
provided with a Fulkerson covering.
\end{thm}
\begin{prf}
We give first a Fulkerson covering of $G_{3}$ in Figure
\ref{Figure:Goldberg3Fulkerson}. The reader will complete easily the
matchings along the $5-$cycles by remarking that these cycles are
incident to $5$ edges with a common label from $1$ to $6$ and to
exactly one edge of each remaining label. We obtain a Fulkerson
covering of $G_{k}$ with odd $k\geq 5$ by inserting a suitable
number of subgraphs isomorphic to the subgraph depicted in Figure
\ref{Figure:MaillonGoldberg} when we cut $G_{3}$ along the dashed
line. The labels of the edges of the two sets of three semi-edges
(left and right) are identical which insures that the process can be
repeated as long as necessary. These labels lead to the perfect
matchings of the Fulkerson covering.
\begin{figure}
\centering 
\subfigure[A Fulkerson covering for $G_3$]{
\includegraphics[scale=0.8]{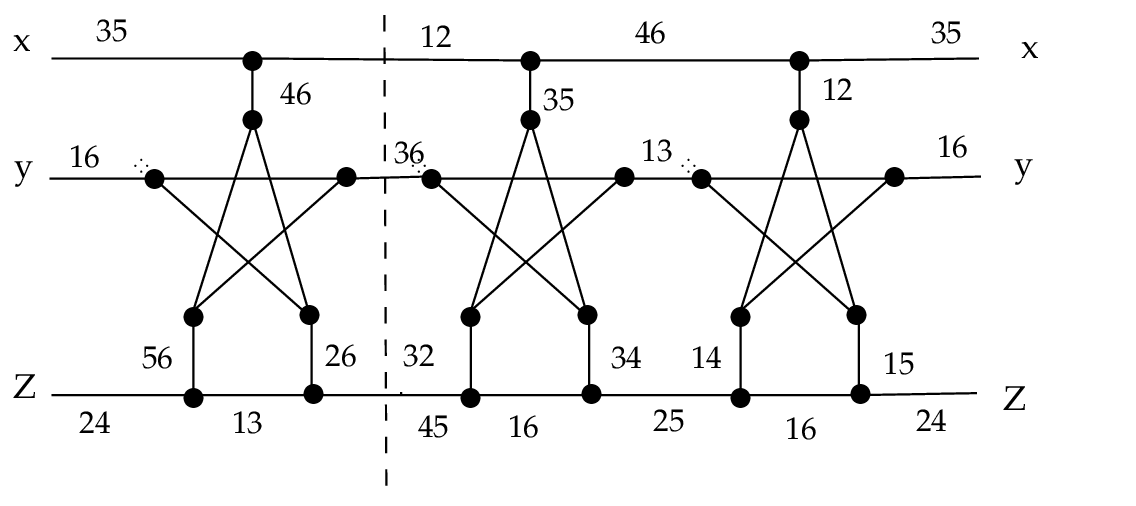}
\label{Figure:Goldberg3Fulkerson}
}
\subfigure[A block for the Goldberg snark]{
\includegraphics[scale=0.8]{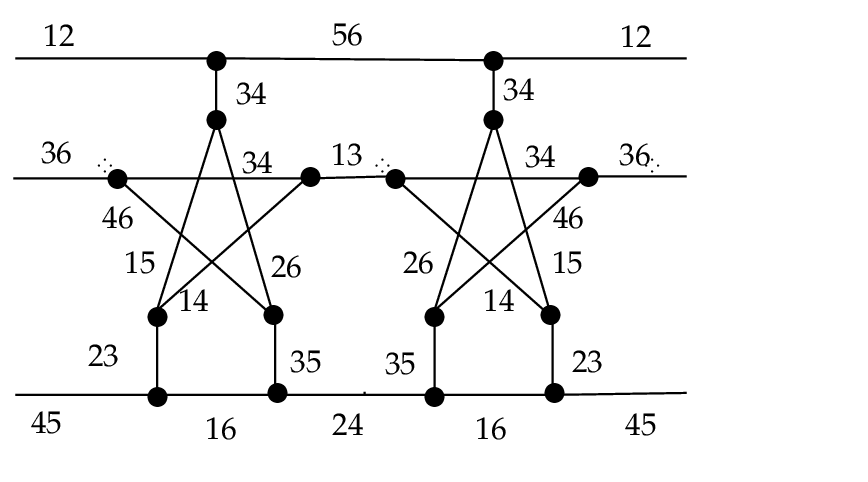}
\label{Figure:MaillonGoldberg}
}
\caption{Fulkerson covering for the Golberg Snarks}
\end{figure}
\end{prf}

\section{A technical tool}

Let $M$ be a perfect matching, a set $A \subseteq E(G)$ is an {\em
$M-$balanced matching} when we can find a perfect matching $M'$
 such that $A=M \cap M'$. Assume that $\mathcal M=\{A,B,C,D\}$  are $4$
 pairwise disjoint $M-$balanced matchings, we shall say that
 $\mathcal M$ is an {\em F-family for $M$} whenever the three following
 conditions are fulfilled:

 \begin{itemize}
   \item [i \label{Item:1}] Every odd cycle of $G \backslash M$ has exactly one
vertex
   incident with one edge of each matching in $\mathcal M$.
   \item [ii \label{Item:2}] Every even cycle of $G \backslash M$ incident with
some matching in $\mathcal M$   contains $4$ vertices such that
   two of them are incident to one matching in $\mathcal M$ while the other are
incident to another matching in $\mathcal
   M$ or the $4$ vertices are incident to the same matching in 
   $\mathcal M$.
   \item [ iii \label{Item:3}] The subgraph induced by $4$ vertices so determined in the previous
items has a matching.
 \end{itemize}
It will be convenient to denote the set of edges described in the
third item by $N$.
\begin{thm} \label{Theorem:TechnicalTool}  Let $G$ be a bridgeless cubic graph
together with
a perfect matching $M$ and an F-family for $M$ $\mathcal M$. Then $G$
can be provided with a Fulkerson covering.
\end{thm}
\begin{prf} Since $A,B,C$ and $D$ are $M-$balanced matchings, we can
find $4$ perfect matchings $M_{A}$, $M_{B}$, $M_{C}$ and $M_{D}$
such that
$$M \cap M_{A}=A \quad M \cap M_{B}=B \quad M \cap M_{C}=C \quad M \cap
M_{D}=D$$

Let $M'=M \setminus \{A,B,C,D\} \cup N$, we will prove that
$\mathcal F=\{M,M_{A},M_{B},M_{C},M_{D},M'\}$ is a Fulkerson
covering of $G$.

\begin{Clm}\label{Claim:NperfectMatching}
$M'$ is a perfect matching
\end{Clm}
\begin{PrfClaim}
The vertices of $G$ which are not incident with some edge in $M
\setminus \{A,B,C,D\}$ are precisely those which are end vertices of
edges in $M_A\cup M_B\cup M_C\cup M_D$. From the definition of an  F-family, the $4$
vertices defined on each cycle of $\{C_{i} | i=1 \ldots k\}$
incident to edges of $\mathcal M$ form a matching with two edges,
which insures that $M'$ is a perfect matching.

\end{PrfClaim}
Let $\mathcal C=\{\Gamma_{i} | i=1 \ldots k\}$ be the set of cycles
of $G \backslash M$ and let $X$ and $Y$ be two distinct members of
$\mathcal M$.
\begin{Clm}\label{Claim:OddCycles}
Let $\Gamma \in \mathcal C$ be an odd cycle. Assume that $X$ and $Y$
have ends $x$ and $y$ on $\Gamma$. Then  $xy$ is the only edge of
$C$ not covered by $M_{X} \cup M_{Y}$
\end{Clm}
\begin{PrfClaim}
Since $M_{X}$ ($M_{Y}$ respectively) is a perfect matching, the
edges of $M_{X}$ ($M_{Y}$ respectively) contained in $\Gamma$
saturate every vertex of $\Gamma$ with the exception of $x$ ($y$
respectively). The result follows.
\end{PrfClaim}

\begin{Clm}\label{Claim:EvenCyclesFirst}
Let $\Gamma \in \mathcal C$ be an even cycle. Assume that $X$ and
$Y$ have ends $x_{1},x_{2}$ and $y_{1},y_{2}$ on $C$ with
$x_{1}y_{1} \in N$ and $x_{2}y_{2} \in N$. Then $x_{1}y_{1}$ and
$x_{2}y_{2}$ are the only edges of $\Gamma$ not covered by $M_{X}
\cup M_{Y}$
\end{Clm}
\begin{PrfClaim} The perfect matching $M_{X}$ must saturate every vertex of
$\Gamma$ with the
exception of $x_{1}$ and $x_{2}$. The same holds with $M_{Y}$ and
$y_{1}$ and $y_{2}$. Since $x_1y_{1}$ and $x_{2}y_{2}$ are edges of
$\Gamma$, these two edges are not covered by $M_{X} \cup M_{Y}$ and we
can easily check that the other edges are covered.

\end{PrfClaim}

\begin{Clm}\label{Claim:EvenCyclesSecond}
Let $\Gamma \in \mathcal C$ be an even cycle. Assume that $X$ and
$Y$ have ends $x_{1},x_{2}$ and $y_{1},y_{2}$ on $C$ with
$x_{1}x_{2} \in N$ and $y_{1}y_{2} \in N$. Then either $x_{1}x_{2}$
and $y_{1}y_{2}$ are the only edges of $\Gamma$ not covered by
$M_{X} \cup M_{Y}$ or $M_{X} \cup M_{Y}$ induces a perfect matching
on $\Gamma$ such that every edge in that perfect matching is covered
by $M_{X}$ and $M_{Y}$ with the exception of $x_{1}x_{2}$ which
belongs to $M_{Y}$ and $y_{1}y_{2}$ which belongs to $M_{X}$.
\end{Clm}
\begin{PrfClaim}The perfect matching $M_{X}$ must saturate every vertex of
$\Gamma$ with the
exception of the two consecutive vertices $x_{1}$ and $x_{2}$. The
same holds with $M_{Y}$ and $y_{1}$ and $y_{2}$.

Let us recall here that, since  $X$ ($Y$ respectively) is a balanced
matching, the paths determined by $x_{1}$ and $x_{2}$ on $\Gamma$
have odd lengths (the paths determined by $y_{1}$ and $y_{2}$
respectively). Two cases may occur.

{\bf case 1:} {\it The two paths obtained on $\Gamma$ by deleting
the edges $x_{1}x_{2}$ and $y_{1}y_{2}$  have odd lengths} We can
check that $M_{X} \cup M_{Y}$ determines a perfect matching on
$\Gamma$ such that every edge in that perfect matching is covered by
$M_{X}$ and $M_{Y}$ with the exception of $x_{1}x_{2}$ which belongs
to $M_{Y}$ and $y_{1}y_{2}$ which belongs to $M_{X}$

{\bf case 2:} {\it The two paths obtained on $\Gamma$ by deleting
$x_{1}x_{2}$ and $y_{1}y_{2}$  have even lengths} We can check that
$M_{X} \cup M_{Y}$ covers every edge of $\Gamma$ with the exception
of $x_{1}x_{2}$ and $y_{1}y_{2}$.

\end{PrfClaim}

\begin{Clm}\label{Claim:EvenCyclesThird}
Let $\Gamma \in \mathcal C$ be an even cycle. Assume that $X$ have
ends $x_{1},x_{2},x_{3}$ and $x_{4}$ on $\Gamma$ with $x_{1}x_{2}
\in N$ and $x_{3}x_{4} \in N$.  Then we can choose a perfect
matching $M_{Y}$ in such a way that $x_{1}x_{2}$ and $x_{3}x_{4}$
are the only edges of $\Gamma$ not covered by $M_{X} \cup M_{Y}$.
\end{Clm}
\begin{PrfClaim}
Since $M_{X}$ is a perfect matching, the edges of $M_{X}$  contained
in $\Gamma$ saturate every vertex of $\Gamma$ with the exception of
$x_{1},x_{2},x_{3}$ and $x_{4}$. Since  $Y$ is not incident to
$\Gamma$ the perfect matching $M_{Y}$ can be chosen in two ways
(taking one of the two perfect matchings contained in this cycle).
We can see easily that we can choose $M_{Y}$ in such a way that
every edge distinct from $x_{1}x_{2}$ and $x_{3}x_{4}$ is covered by
$M_{X}$ or $M_{Y}$.
\end{PrfClaim}

Since $\{A,B,C,D,M'\cap M\}$ is a partition of $M$, each edge of $M$
is covered twice by some perfect matchings of $\mathcal F$.

Let $\Gamma \in \mathcal C$ be an odd cycle, each edge of $\Gamma$
distinct from the two edges of $N$ (Claim \ref{Claim:OddCycles}) is
covered twice by some perfect matchings of $\mathcal F$. The two
edges of $N$  are covered by exactly one perfect matching belonging
to $\{M_{A},M_{B},M_{C},M_{D}\}$ and by the perfect matching $M'$.
Hence every edge of $\Gamma$ is covered twice by $\mathcal F$.

Let $\Gamma \in \mathcal C$ be an even cycle. Assume first that $4$
vertices of $\Gamma$ are ends of some edges in $A$ while no other
set of $\mathcal M$ is incident with $\Gamma$. From Claim
\ref{Claim:EvenCyclesThird} we can choose $M_{B}$ in such a way that
 every edge distinct from the two edges of $N$  is
covered by  $M_{A}$ or $M_{B}$. We can then choose $M_{C}$ in such a
way that one of the two edges of $N$ belongs to $M_{C}$. Finally, we
can choose $M_{D}$ in order to cover the other edge of $N$. Each
edge of $\Gamma$ distinct from the two edges of $N$ (Claim
 \ref{Claim:EvenCyclesThird}) is
covered twice by some perfect matchings of $\mathcal F$. The two
edges of $N$  are covered by exactly one perfect matching belonging
to $\{M_{A},M_{B},M_{C},M_{D}\}$ and by the perfect matching $M'$.
Hence every edge of $\Gamma$ is covered twice by $\mathcal F$.

Assume now that $2$ vertices of $\Gamma$ are ends of some edges in
$A$ (say $a_{1}$ and $a_{2}$) and $2$ other vertices are ends of
some edges in $B$ (say $b_{1}$ and $b_{2}$).

 {\bf case 1:} $a_{1}b_{1} \in N$ and $a_{2}b_{2} \in N$. We can choose $M_{C}$
and $M_{D}$ in order to cover every edge of $\Gamma$. From Claim
\ref{Claim:EvenCyclesFirst} every edge of $\Gamma$ is covered by
$M_{A} \cup M_{B}$ with the exception of $a_{1}b_{1}$ and
$a_{2}b_{2}$. Hence every edge of $\Gamma$ is covered twice by
$M_{A} \cup M_{B} \cup M_{C} \cup M_{D}$ while $a_{1}b_{1}$ and
$a_{2}b_{2}$ are covered twice by $M_{C} \cup M_{D} \cup M^{'}$
Hence every edge of $\Gamma$ is covered twice by $\mathcal F$.

 {\bf case 2:} $a_{1}a_{2} \in N$ and $b_{1}b_{2} \in N$.
Assume that $a_{1}a_{2}$ and $b_{1}b_{2}$ are the only edges of
$\Gamma$ not covered by $M_{A} \cup M_{B}$ (Claim
\ref{Claim:EvenCyclesSecond}). Then we can choose $M_{C}$ and $M_D$
in such a way that every edge of $\Gamma$ is covered by $M_{C} \cup
M_{D}$. In that case every edge of $\Gamma$ is covered twice by
$M_{A} \cup M_{B} \cup M_{C} \cup M_{D}$ with the exception of
$a_{1}a_{2}$ and $b_{1}b_{2}$ which are covered twice by $M_{C} \cup
M_{D} \cup M^{'}$.

Assume now that  $M_{A} \cup M_{B}$ induces a perfect matching on
$\Gamma$ where $a_{1}a_{2} \in M_{B}$ and $b_{1}b_{2} \in M_{A}$
while the other edges of this perfect matchings are in $M_{A} \cap
M_{B}$ (Claim \ref{Claim:EvenCyclesSecond}). Then we can choose
$M_{C}$ and $M_{D}$ such that every edge of $\Gamma$ not contained
in $M_{A} \cup M_{B}$ is covered twice by $M_{C} \cup M_{D}$ ($M_{C}
\cup M_{D}$ induces a perfect matching on $\Gamma$). hence every
edge of $\Gamma$ is covered twice by $M_{C} \cup M_{D}$ or by $M_{A}
\cup M_{B}$ with the exception of $a_{1}a_{2}$ which is covered
twice by $M_{B} \cup M^{'}$ and $b_{1}b_{2}$ which is covered twice
by $M_{A} \cup M^{'}$.

Finally, assume that $\Gamma$ has no vertex as end of some edge in
$\mathcal M$. Then we can choose easily $M_{A}, M_{B}, M_{C}$ and
$M_{D}$ such that every edge of $\Gamma$ is covered twice by
$M_{A}\cup M_{B}\cup M_{C}\cup M_{D}$

Hence $\mathcal F$ is a Fulkerson covering of $G$.
\end{prf}
\begin{rem} \label{Rem:TousDistincts}
Observe that the matchings of the Fulkerson covering described in the above
proof are all distinct.
\end{rem}
\subsection{Dot products which preserve an F-family}
In \cite{Isa75} Isaacs defined the {\em dot product} operation in order to
describe infinites families of non trivial snarks.

Let $G_{1}, G_{2}$ be two bridgeless cubic graphs and
$e_{1}=u_{1}v_{1}$, $e_{2}=u_{2}v_{2} \in E(G_{1})$ and
$e_{3}=x_{1}x_{2} \in E(G_{2})$ with
$N_{G_{2}}(x_{1})=\{y_{1},y_{2},x_{2}\}$ and
$N_{G_{2}}(x_{2})=\{z_{1},z_{2},x_{1}\}$.

\begin{figure}
\begin{center}
\includegraphics[scale=0.8]{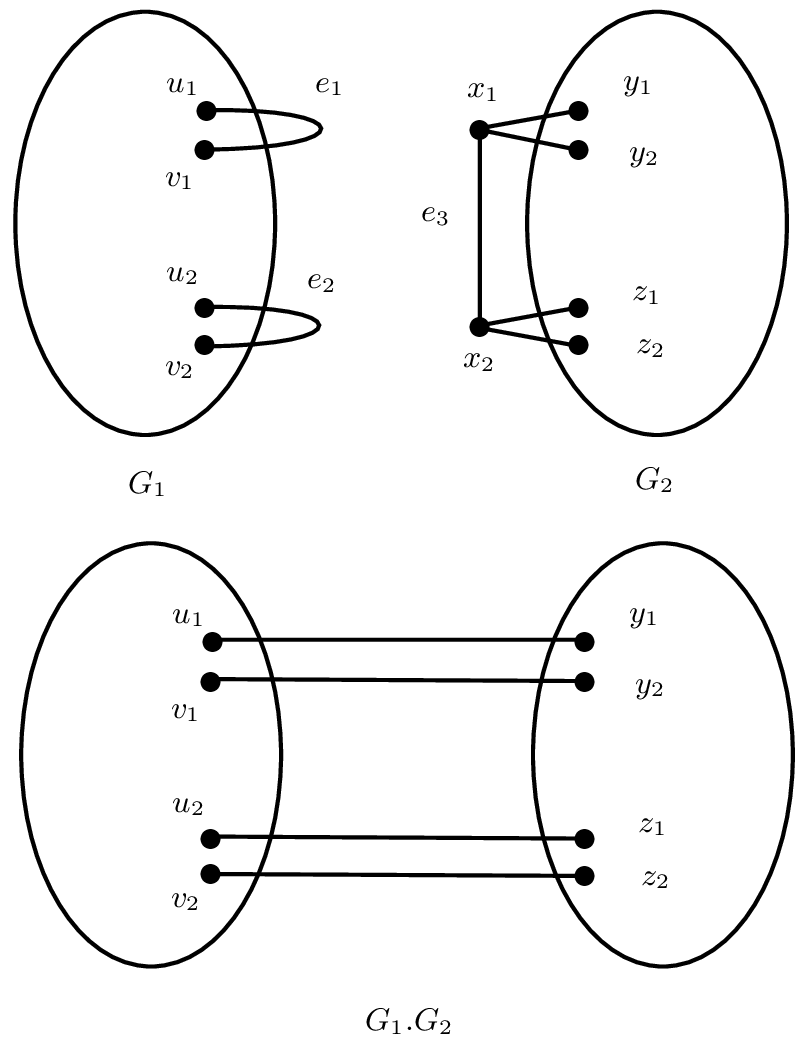}
\end{center}
\caption{The dot product operation\label{Fig:DotProduct}} 
\end{figure}

The {\em dot product} of  $G_1$ and $G_2$, denoted by $G_1 \cdot
G_2$
is the bridgless cubic graph $G$ defined
as  (see Figure \ref{Fig:DotProduct})~:
$$G=[G_{1} \backslash \{e_{1},e_{2}\}] \cup [G_{2} \backslash
\{x_{1},x_{2}\}] \cup
\{u_{1}y_{1},v_{1}y_{2},u_{2}z_{1},v_{2}z_{2}\}$$

It is well known that the dot product of two snarks remains to be a
snark. It must be pointed out that in general the dot product
operation does not permit to extend a Fulkerson covering, in other
words, whenever $G_1$ and $G_2$ are snarks together with a Fulkerson
covering, we do not know how to get a Fulkerson covering for $G_1
\cdot G_2$.

However, in some cases, the dot product operation can preserve, in
some sense, an $F-$family, leading thus to a Fulkerson covering of the
resulting graph.
\begin{prop}\label{Proposition:DotProduct_1}
Let $M_{1}$ be a perfect matching of a snark $G_1$ such that
$G_1\backslash M_1$ contains only two (odd) cycles, namely $C$ and
$C'$.  Let $ab$ be an edge of $C$ and $a'b'$ be an edge of $C'$.

Let $M_2$  be a perfect matching of a snark $G_2$ where  $\{A,B,C,D\}$ is an $F-$family for $M_2$.
Let $xy$ be an edge of $M_{2} \backslash \{A \cup B \cup C \cup
D\}$, with $x$ and $y$  vertices of two distinct odd cycles of
$G_2\backslash M_2$.

Then $\{A,B,C,D\}$ is an $F$-family for the perfect matching $M$ of
$G=G_1 \cdot G_2$ with $M=M_1\cup M_{2}\backslash \{xy\}$.
\end{prop}
\begin{prf}
Obvious by the definition of the $F-$family and the  construction of
the graph resulting of the dot product.
\end{prf}
\begin{prop}\label{Proposition:DotProduct_2}
Let $M_1$  be a perfect matching of a snark $G_1$ where  $\{A,B,C,D\}$ is an $F-$family for $M_1$.
Let $xy$ and $zt$ be two edges of $E(G_{1}) \backslash M_{1}$  not
contained in $N$.

Let $M_{2}$ be a perfect matching of a snark $G_2$ such that
$G_2\backslash M_2$ contains only two (odd) cycles, namely $C$ and
$C'$. Let $xy \in M_{2}$, with $x \in V(C)$ and $y \in V(C')$.

Then $\{A,B,C,D\}$ is an $F$-family for the perfect matching $M$ of
$G=G_1 \cdot G_2$ with $M=M_1\cup M_{2}\backslash \{xy\}$.
\end{prop}
\begin{prf}
Obvious by the definition of the $F-$family and the  construction of
the graph resulting of the dot product.
\end{prf}

We remark that the graphs obtained via Propositions
\ref{Proposition:DotProduct_1} and \ref{Proposition:DotProduct_2}
can be provided with a Fulkerson covering by Theorem
\ref{Theorem:TechnicalTool}. 

The dot product operations described in Propositions \ref{Proposition:DotProduct_1} and \ref{Proposition:DotProduct_2} will be said to {\em preserve} the F-family.

\section{Applications}
\begin{figure}
\includegraphics[scale=1]{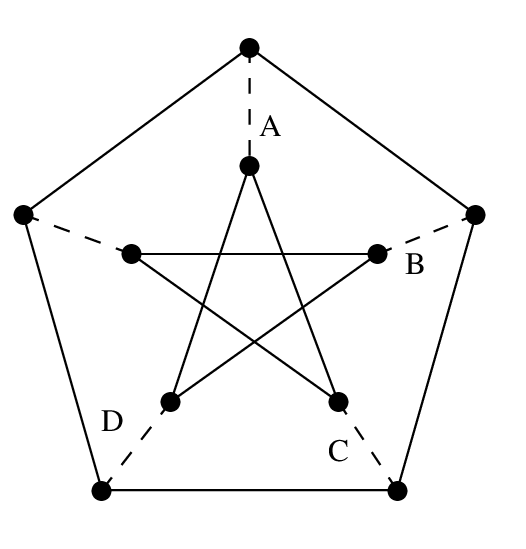}
 \caption{An F-family $\{A,B,C,D\}$ for the Petersen graph.} \label{Fig:FfamilyDansPetersen}
\end{figure}
\begin{figure}
\subfigure[An F-family $\{A,B,C,D\}$ for the flower snark $J_5$.]{
  \includegraphics[scale=1]{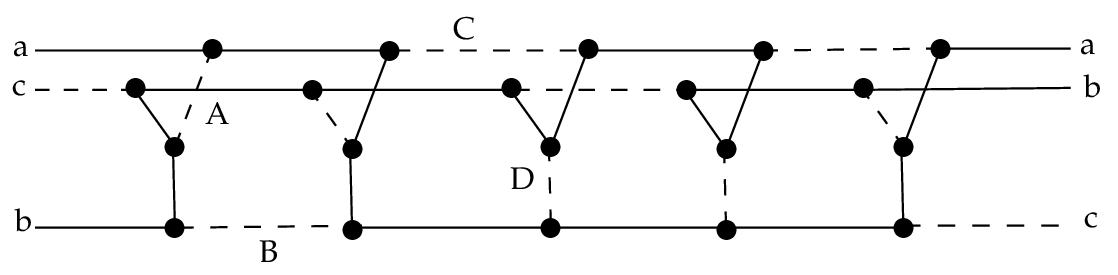}\label{subfig:FfamilyDansJ5}}

\subfigure[Two more Claws.]{
  \includegraphics[scale=1]{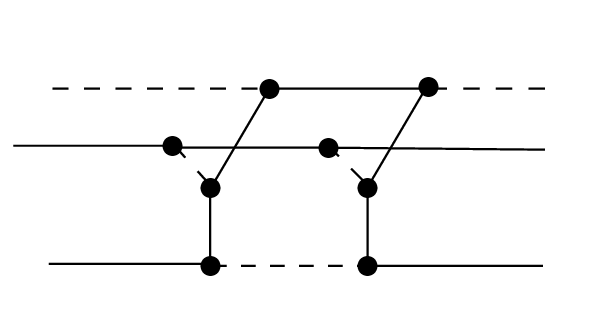}\label{subfig:TwoMoreClaws}
}
  \caption{An F-family $\{A,B,C,D\}$ for the flower snark $J_k$.} 
\label{Fig:FfamilyDansJk}
\end{figure}
\subsection{Fulkerson coverings, more examples}
Figures \ref{Fig:FfamilyDansPetersen} and \ref{subfig:FfamilyDansJ5} show
that the Petersen Graph as well as the flower snark $J_5$ have oddness $2$ and
have an F-family (the dashed edges denote the related perfect matching).

Moreover, as shown in Figure \ref{subfig:TwoMoreClaws} the F-family of $J_5$ can be extended by induction to all the $J_k$'s ($k$ odd).

Thus, following the above Propositions we can define a sequence $(G_n)_{n\in
\mathbb{N}}$ of cubic graphs as follows ~:
\begin{itemize}
 \item Let $G_0$ be the Petersen graph or the flower snark $J_k$ ($k>3$, $k$ odd).
 \item For $n\in \mathbb{N}^*$, $G_n=G_{n-1} . G$ where $G$ is either the
Petersen graph or the flower snark $J_k$ ($k>3$, $k$ odd) and the dot product operation preserves
the F-family.
\end{itemize}
\begin{figure}
\includegraphics[scale=0.6]{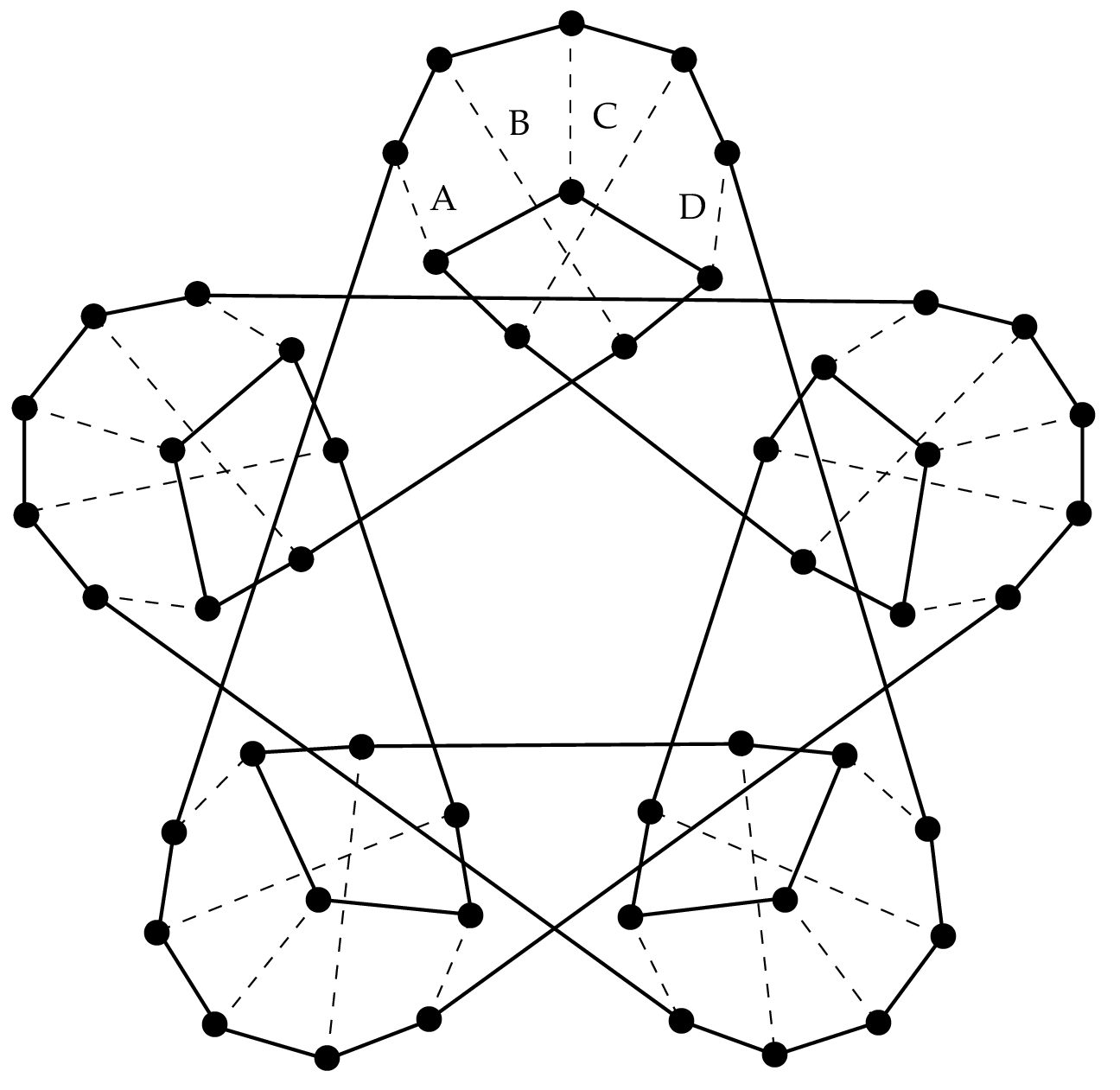}
\caption{An F-family $\{A,B,C,D\}$ for the Szekeres Snark.} \label{Fig:FfamilyDansSzekeres}
\end{figure}
\begin{figure}
 \subfigure[Blanu\v{s}a snark of type $1$]{
  \includegraphics[scale=0.7]{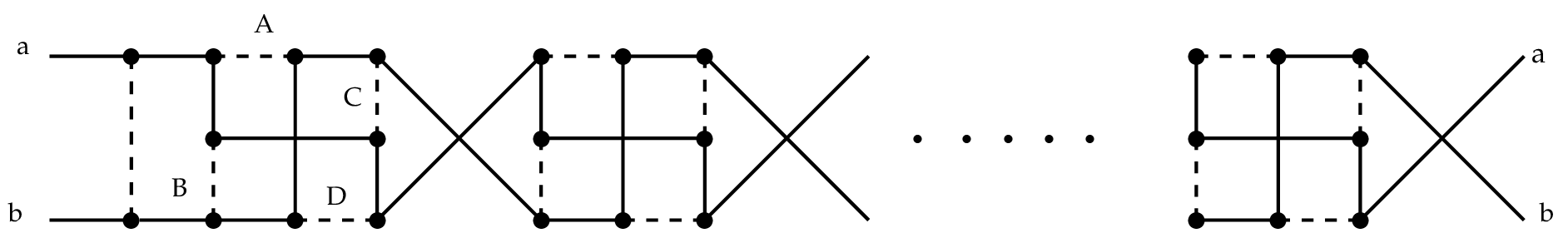}\label{subfig:FfamilyDansBlanusa1}
 }
\subfigure[Blanu\v{s}a snark of type $2$]{
  \includegraphics[scale=0.7]{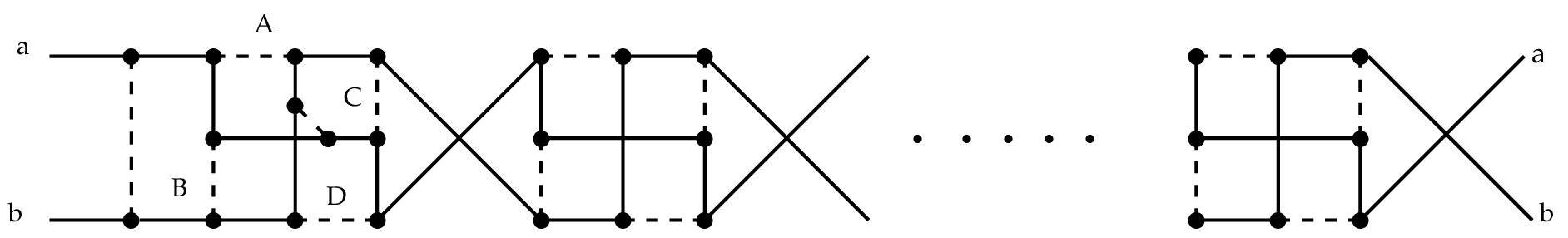}\label{subfig:FfamilyDansBlanusa2}
 }
\caption{An F-family $\{A,B,C,D\}$ for the Generalized Blanu\v{s}a snarks} \label{Figure:GeneralizedBlanusa}
\end{figure}

As a matter of fact this sequence of iterated dot products of the Petersen graph
and/or the flower snark $J_k$ forms a family of exponentially many snarks
including the Szekeres Snark (see Figure \ref{Fig:FfamilyDansSzekeres}) as well as the two types of generalized Blanu\v{s}a snarks proposed by Watkins in \cite{Wat89} (see Figure \ref{Figure:GeneralizedBlanusa}). 

The family obtained when reducing the possible values of $k$ to $k=5$  has already been defined by Skupie\'n in \cite{Sku89}, in order to provide a family
of hypohamiltonian snarks in using the so-called {\em Flip-flop construction}
introduced by Chv\'atal in \cite{Chv73}.

As far as we know there is no Fulkerson family for the Golberg snark.

\subsection{Graphs with a $2$-factor of $C_5$'s.}
Let $G$ be a bridgeless cubic graph having a $2-$factor where each
cycle is isomorphic to a chordless $C_{5}$. We denote by $G^{*}$ the
multigraph obtained from $G$ by shrinking each $C_{5}$ of this
$2-$factor in a single vertex. The graph $G^{*}$ is $5-$regular and
we can easily check that it is bridgeless.

\begin{thm} \label{Theorem:2FactorC5}  Let $G$ be a bridgeless cubic graph
having a $2-$factor
of chordless $C_{5}$. Assume that  $G^{*}$
 has chromatic index $5$. Then $G$ can be provided with a Fulkerson covering.
\end{thm}
\begin{prf} Let $M$ be the perfect matching complementary of the $2-$factor of
$C_{5}$.
Let $\{A,B,C,D,E\}$ be a $5-$edge colouring of $G^{*}$. Each colour
corresponds to a matching of $G$ (let us denote these matchings by
$A,B,C,D$ and $E$). Then it is an easy task to see that $\mathcal
M=\{A,B,C,D\}$ is an F-family for $M$ and the result follows from
Theorem \ref{Theorem:TechnicalTool}.
\end{prf}

\begin{thm} \label{Cor:2FactorC5Biparti}  Let $G$ be a bridgeless cubic graph
having a $2-$factor of
 chordless $C_{5}$. Assume that  $G^{*}$
 is bipartite. Then $G$ can be provided with a Fulkerson covering.
\end{thm}
\begin{prf}
It is well known, in that case, the chromatic index of $G^{*}$ is
$5$. the result follows from Theorem \ref{Theorem:2FactorC5}.
\end{prf}

Remark that, when considering the Petersen graph  $P$, the graph
associated $P^{*}$ is reduced to two vertices and is thus bipartite.

We can construct  cubic graphs with chromatic index $4$  which are cyclically $4$- edge connected ({\em
snarks} in the literature)
and having a $2$-factor of $C_5$'s. Indeed, let $G$ be cyclically
$4$-edge connected snark of size $n$ and $M$ be a perfect matching
of $G$, $M=\{x_iy_i | i=1\ldots \frac{n}{2}\}$. Let $G_1\ldots
G_{\frac{n}{2}}$ be $\frac{n}{2}$
 cyclically $4$-edge connected snarks (each of them having a
$2$-factor of $C_{5}$). For each $G_i$ ($i=1\ldots \frac{n}{2}$) we
consider two edges $e_i^1$ and $e_i^2$ of the perfect
matching which is the complement of the $2-$factor.

We construct then a new cyclically $4$-edge connected snark $H$  by
applying the dot-product operation on $\{e_i^1,e_i^2\}$ and the edge $x_iy_i$ ($i=1\ldots
\frac{n}{2}$). We remark that the vertices of $G$ vanish in the
operation and the resulting graph $H$ has a $2$ factor of $C_5$,
 which is the union of the $2-$factors of $C_{5}$ of the
$G_{i}$. Unfortunately, when considering the graph $H^{*}$, derived
from $H$, we cannot insure, in general, that $H^{*}$ is $5-$edge
colourable in order to apply Theorem \ref{Theorem:2FactorC5} and
obtain hence a Fulkerson covering of $H$.

An interesting case is obtained when, in the above construction of
$H$, each graph $G_{i}$ is isomorphic to the Petersen graph. Indeed,
the $2-$factor of $C_5$'s obtained then is such that we can find a
partition of the vertex set of $H$ in sets of $2$ $C_{5}$ joined
by $3$ edges. These sets lead to pairs of vertices of $H^{*}$ joined
by three parallel edges. We can thus see $H^{*}$ as a cubic graph
where a perfect matching is taken $3$ times. Let us denote by
$\tilde{H}$ this cubic graph (by the way $\tilde{H}$ is
$3$-connected). It is an easy task to see that, when $\tilde{H}$ is
$3-$edge colourable, $H^{*}$ is $5-$edge colourable and hence,
Theorem \ref{Theorem:2FactorC5} can be applied.

\begin{figure}
\includegraphics[scale=1]{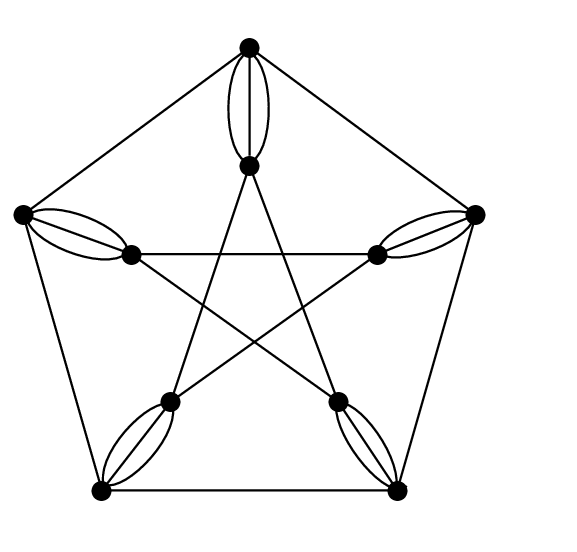}
 \caption{$H^{*}$ isomorphic to $\mathcal
P(3)$} \label{Figure:UnslicableP3}
\end{figure}

Let us consider by example the graph $H$ obtained with $5$ copies of
the Petersen graph following the above construction (let us remark
that the graph  $G$ involved in our construction  must be isomorphic
also to the Petersen graph). This graph is a snark on $50$ vertices.
Since $\tilde{H}$ is a bridgeless cubic graph, the only case for
which we cannot say whether $H$ has a Fulkerson covering occurs when
$\tilde{H}$ is isomorphic to the Petersen graph and, hence $H^{*}$
is isomorphic to the {\em unslicable} graph $\mathcal P(3)$
described by Rizzi \cite{Riz99} (see Figure
\ref{Figure:UnslicableP3}). As a matter of fact we do not know if it is possible to construct a graph $H$ as described above such that $H^*$ is isomorphic to the graph $\mathcal P(3)$.

By the way, we do not know example of cyclically $5$-edge connected
snarks (excepted the Petersen graph)  with a $2$-factor of induced
cycles of length $5$. We have proposed in \cite{FouVan09a} the
following problem.

\begin{prob} Is there any $5$-edge connected snark distinct from the Petersen
graph
with a $2$-factor of $C_5$'s~?
\end{prob}

\section{On proper Fulkerson covering\label{Section:FulkersonProperCovering}}

As noticed in the introduction, when a cubic graph is $3-$edge
colourable, we can find a Fulkerson covering by using a $3-$edge
colouring and considering each colour twice.
\begin{prop} \label{Proposition:ProperIndex4}
Let $G$ be a bridgeless cubic graph with chromatic index $4$. Assume
that $G$ has a Fulkerson covering $\mathcal F =\{M_{1},
M_{2},M_{3},M_{4},M_{5},M_{6}\}$ of its edge set. Then the $6$
perfect matchings are distinct.
\end{prop}
\begin{prf}Assume, without loss of generality that $M_{1}=M_{2}$.
Since each edge is contained in exactly $2$ perfect matchings of
$\mathcal F$, we must have $M_{3} \cap M_{1}= \emptyset$. Hence $G$
is $3-$edge colourable, a contradiction.
\end{prf}

Let us say that a Fulkerson covering is {\em proper} whenever the
$6$ perfect matchings involved in this covering are distinct. An
interesting question is thus to determine which cubic bridgeless
graph have a proper Fulkerson covering.

A $3$-edge colourable graph is said to be {\em bi-hamiltonian} whenever in any
$3$-edge colouring, there are at least two colours whose removing leads to an
hamiltonian $2$-factor.

\begin{prop} \label{Proposition:PropernonHamiltonian}
Let $G$ be a bridgeless $3$-edge colourable cubic graph which is not bi-hamiltonian. Then $G$ has a
proper Fulkerson covering.
\end{prop}
\begin{prf}
Let $\Phi \quad E(G) \rightarrow \{\alpha,\beta,\gamma\}$ be a  $3-$edge
colouring of $G$. When $x$ and $y$ are colours in
$\{\alpha,\beta,\gamma\}$,  $ \Phi(x, y)$ denotes the set of disjoint
even cycles induced by the two colours $x$ and $y$.
\newline Since the graph $G$ is not bi-hamiltonian we may assume that the
$2$-factors $\Phi(\alpha,\beta)$ and $\Phi(\beta,\gamma)$ are not hamiltonian
cycles.
Let $C$ be a
cycle in $\Phi(\alpha,\beta)$, we get a new $3-$edge colouring
$\Phi^{'}$ by exchanging the two colours $\alpha$ and $\beta$ along
$C$. We get hence a partition of $E(G)$ into $3$ perfect matching
$\alpha^{'},\beta^{'}$ and $\gamma$. In the same way, when
considering a cycle $D$ in $\Phi(\beta,\gamma)$, we get a new
$3-$edge colouring $\Phi^{''}$ of $G$ by exchanging $\beta$ and
$\gamma$ along $D$. Let $\alpha$,$\beta^{''}$ and $\gamma^{''}$ be
the $3$ perfect matchings so obtained.

Since we have two distinct $3-$edge colourings of $G$,
$\Phi^{'}$and $\Phi^{''}$, the set of $6$ perfect matchings so
involved
$\{\alpha,\alpha^{'},\beta^{'},\beta^{''},\gamma,\gamma^{''}\}$ is a
Fulkerson covering. It remains to show that this set is actually a
proper Fulkerson covering.

The exchange operated in
order to get $\Phi^{'}$ involve some edges in $\alpha$ and some
edges in $\beta$ (those which are on $C_{1}$) while the other edges
keep their colour. In the same way,  the exchange operated in order
to get $\Phi^{''}$ involve some edges in $\beta$ and some edges in
$\gamma$ (those which are on $D_{1}$) while the other edges keep
their colour.

The $3$ perfect matchings of $\Phi^{'}$ ($\alpha^{'},\beta^{'}$ and
$\gamma$ ) are pairwise disjoint as well as those of $\Phi^{''}$
($\alpha,\beta^{''}$ and $\gamma^{''}$).  We have $\alpha \not =
\alpha^{'}$ since $\alpha^{'}$ contains some edges of $\beta$. We
have $\alpha \cap \beta^{''}=\emptyset$ and $\alpha \cap
\gamma^{''}=\emptyset$ since we have exchanged $\beta$ and $\gamma$
in order to obtain $\beta^{''}$ and $\gamma^{''}$. We have
$\beta^{'} \not = \beta^{''}$ since $\beta^{'}$ contains some edges
of $\alpha$ while $\beta^{''}$ contains some edges of $\gamma$. We
have $\beta^{'} \not = \gamma^{''}$ since $\beta^{'}$ contains some
edges of $\alpha$ and $\gamma^{''}$ contains only edges in $\beta$
or in $\gamma$. We have $\gamma \not = \gamma^{''}$ since
$\gamma^{''}$ contains some edges of $\beta$.

Hence
$\{\alpha,\alpha^{'},\beta^{'},\beta^{''},\gamma,\gamma^{''}\}$ is a
proper Fulkerson covering.

\end{prf}

The {\em theta graph }(2 vertices joined by 3 edges), $K_{4}$,
$K_{3,3}$ are examples of small bridgeless cubic graph without
proper Fulkerson covering. The infinite family of bridgeless cubic
bi-hamiltonian graphs obtained by doubling the edges of a perfect matching of an
even cycle has no proper Fulkerson covering.
On the other hand, we can provide a bi-hamiltonian graph together with a proper
Fulkerson covering. Consider for example the graph $G$ on $10$ vertices which
have a $2$ factor of $C_5$'s, namely $abcde$ and $12345$ with the additional
edges edges $a2$, $b4$, $c3$, $d5$ and $e1$, it is not difficult to check that
this graph is bi-hamiltonian. Moreover since the following four balanced
matchings $\{a2\}$, $\{b4\}$, $\{c3\}$ and $\{d5\}$ form an F-family for the
perfect matching $\{a2,b4,c3,d5,e1\}$, due to Theorem
\ref{Theorem:TechnicalTool} and Remark \ref{Rem:TousDistincts}, the graph $G$
has a proper Fulkerson covering.

A challenging problem is thus
to characterize those bridgeless cubic graphs having a proper Fulkerson
covering.

{\bf Acknownledgement.} The authors are gratefull to Professor Skupie\'n for his
helpfull comments on the Flip-flop construction.
\bibliographystyle{amsplain}
\bibliography{BibliographieBergeFulkerson}

\providecommand{\bysame}{\leavevmode\hbox to3em{\hrulefill}\thinspace}
\providecommand{\MR}{\relax\ifhmode\unskip\space\fi MR }
\providecommand{\MRhref}[2]{%
  \href{http://www.ams.org/mathscinet-getitem?mr=#1}{#2}
}
\providecommand{\href}[2]{#2}
\begin{thebibliography}{10}

\bibitem{Chv73}
V.~Chv\'atal, \emph{Flip-flops on hypohamiltonian graphs}, Canad. Math. Bull.
  \textbf{16} (1973), 33--41.

\bibitem{FanRas}
G.~Fan and A.~Raspaud, \emph{Fulkerson's conjecture and circuit covers}, J.
  Comb. Theory Ser. B \textbf{61} (1994), 133--138.

\bibitem{FouVan09a}
J.L. Fouquet and J.M. Vanherpe, \emph{On parsimonious edge-colouring of graphs
  with maximum degree three}, Tech. report, LIFO, april 2009.

\bibitem{Ful71}
D.R. Fulkerson, \emph{Blocking and anti-blocking pairs of polyhedra}, Math.
  Programming \textbf{1} (1971), no.~69, 168--194.

\bibitem{Gol81}
M.K. Goldberg, \emph{Construction of class 2 graphs with maximum vertex degree
  3}, J. Combin. Theory B \textbf{31} (1981), 282--291.

\bibitem{Isa75}
R.~Isaacs, \emph{Infinite families of non-trivial trivalent graphs which are
  not {Tait} colorable}, Am. Math. Monthly \textbf{82} (1975), 221--239.

\bibitem{Riz99}
R.~Rizzi, \emph{Indecomposable $r$-graphs and some other counterexamples},
  Journal of Graph Theory \textbf{32} (1999), no.~1, 1--15.

\bibitem{Sku89}
Z.~Skupie\'n, \emph{Exponentially many hypohamiltonian graphs}, Graphs,
  Hypergraphs and Matroids III (M.~Borowiecky and Z.~Skupie\'n, eds.), Proc.
  Conf.Kalsk 1988, Higher College of Engeniering, Zielona G\'ora, 1989, New
  York, pp.~123--132.

\bibitem{Wat89}
J.~J. Watkins, \emph{Snarks}, Graph theory and its applications: East and West
  (Jinan,1986), Ann. New York Acad. Sci., vol. 576, New York Acad. Sci., 1989,
  New York, pp.~606--622.

\bibitem{HaoNiuWanZhaZha2009}
Rongxia Hao{,} Jianbing Niu{,} Xiaofeng Wang{,} Cun-Quan Zhang and Taoye Zhang,
  \emph{A note on {B}erge-{F}ulkerson coloring}, Discrete Mathematics
  \textbf{309} (2009), no.~13, 4235--4240.

\end{thebibliography}
\end{document}